\documentclass{mn2e}

\usepackage[dvips]{graphicx}

\begin{document}

\title[DLAs in Hierarchical Clustering Scenarios]{Building Blocks in 
Hierarchical Clustering Scenarios and their Connection with 
Damped Ly$\alpha$ Systems}

\author[Cora et al.]{Sof\'{\i}a A. Cora,$^{1,2,5}$, Patricia B. Tissera, $^{2,3}$,
Diego G. Lambas,$^{2,4,6}$  and Mirta B. Mosconi,$^4$   \\
$^1$ Facultad de Ciencias Astron\'omicas y Geof\'{\i}sicas de la Universidad 
Nacional de
La Plata, Argentina.\\
$^2$ Consejo Nacional de Investigaciones Cient\'{\i}ficas
y T\'ecnicas, Argentina.\\
$^3$Instituto de Astronom\'{\i}a.
y F\'{\i}sica del Espacio, Argentina.\\
$^4$ Observatorio Astron\'omico
de la Universidad Nacional de C\'ordoba,  Argentina.\\
$^5$ Postdoctoral Fellow of Fundaci\'on Antorchas at Max-Planck Institute for
Astrophysics, Germany.\\
$^6$ John Simon Guggenheim Fellow.
}

\maketitle

\begin{abstract}
We carried out a comprehensive analysis of the chemical properties
of the interstellar medium (ISM) and the stellar population (SP) of current
normal galaxies and their progenitors in a hierarchical
clustering scenario. We compared the results with observations
of Damped Lyman-$\alpha$ systems (DLAs) under the hypothesis that,
at least, part of the  observed DLAs could originate in the building
blocks of today normal galaxies.
We used a hydrodynamical cosmological code which includes star formation and
chemical enrichment. Galaxy-like objects are identified at $z=0$
and then followed back in time. Random line-of-sights (LOSs) are drawn
through these structures in order to mimic Damped Lyman $\alpha$
systems. We then analysed the chemical properties of the
ISM and SP along the LOSs.
We found that the progenitors of current
galaxies in the field with  mean $L <0.5 L^* $ and virial
circular velocity of $100-250 \ {\rm km/sec}$ could be 
the associated DLA galaxies. For these systems we detected a trend for  $\langle L/L^* \rangle$ 
to increase 
with redshift. 
We found moderate metallicity evolution
for the [Zn/H], [Fe/H] and [Si/H]. However, when we applied
the observational filter suggested by Boiss\'e et al. (1998) in
order to restrict the sample to the observed limits in densities and
metallicities, we found mild evolution consistent with observational
results that include dust corrections.
 The [Si/Fe] and [S/Fe] show weak $\alpha$-enhancement
in agreement with observations corrected by dust depletion.
We found the  ${\rm \alpha/Fe}$ in the ISM and SP to have more
homogeneous abundances than the [Fe/H] and [Zn/H] ones.
In our models, the global metallicity evolution is driven 
by the high metallicity
and high column density simulated DLAs which have low
impact parameters ($b < 5$ kpc) and SPs with more than $10^{8} {\rm M_{\odot}}$.
Our results suggest that  geometrical effects could be
 the  mechanism responsible for the non-detectability of high
metallicity and high column density  DLAs.
We found sub-DLAs to map preferentially the outskirts of the
simulated DLA galaxies. Hence, they can contribute to the study of 
the metallicity
of the galactic structure as a function of redshift.
An analysis of the metallicity content of the ISMs and SPs
of the galaxy-like objects as a function of redshift show the formation
of a central stellar mass concentration with nearly solar metallicity
at all redshift  while
stars in the outer parts of these objects have lower metallicities.
The gas content gets enriched progressively with redshift and at all
radii.
The abundance properties of the galaxy-like objects and the simulated
DLAs are the results of the contribution of supernovae type Ia and
type II  and gas infall
from the dark matter haloes with a timing settled by their
particular history of evolution in a hierarchical clustering scenario.
Our results suggest that the mild evolution detected in the observations could
arise from  a conspiracy of all these processes. 
 
\end{abstract}


\section{Introduction}

Our understanding of the Universe  has
improved dramatically in the last decades due to the outstanding discoveries
made by large orbiting and ground-based telescopes. 
In particular, Damped Ly$\alpha$ absorptions identified  in
the spectra of distant quasars  provide clues on the chemical
and kinematic properties of systems at different redshifts.
However, the nature of the host  
of the absorbing 
 neutral hydrogen (${\rm H_{I}}$) 
clouds (hereafter DLA galaxies)  remains unclear. At low
redshift,
identified DLAs are associated with galaxies of different morphologies with a 
preference for dwarf and low surface brigthness ones (Le Brun et al. 1997; Rao \& Turnsheck 2000), while at higher $z$, there
is no clear evidence of the morphological characteristics of the DLA galaxies.

Different models have been proposed to explain the kinematics of DLAs.
In particular, Haehnelt et al. (1998), by using
hydrodynamical simulations,  have found that the features in the
velocity 
 width 
distributions  of these systems could be
reproduced by the substructure in Cold Dark Matter models 
(see also MacDonald \& Miralda-Escud\'e 1999).
Other mechanisms have been suggested as responsible for producing such
velocity distributions. Among them, Nulsen, Barcons \& Fabian (1998)
claimed that massive out-flows from dwarf galaxies could account
for the majority of DLA systems in their semi-analytical models for galaxy
formation (see also Schaye 2001a).

>From the point of view of chemical evolution, abundance ratios suggest that DLAs are
young systems with  a low metallicity content which seems not to evolve
significantly from very high $z$ to present days (e.g., Prochaska \& Wolfe 2002). However,
 several biasing factors may be affecting these 
conclusions.  Dust depletion and obscuration are
a main source of uncertainties for the determination of  some element abundances
 (e.g., Vladilo 1998). 
However, the estimations of the amount of depletion and obscuration by dust
for each element is a complex task which have produced different results  (Vladilo 2002; Prochaska \& Wolfe 2002).
Another important effect is that DLAs with  high metallicity and high column
density
 are missing from the data (e.g., Boiss\'e et al. 1998; Hou et al. 2001).
 Traditionally, the lack
of these systems has been interpreted as a bias produced by
dust obscuration. 
However, first analysis of DLAs identified from a radio
selected QSO survey 
by Ellison et al. (2001) could suggest  that such sort of DLAs might not
be present at all in nature since the selection of this sample should not
be dust biased.  Howerver, a larger database is needed before reaching
a solid conclusion.

The chemical properties  of the absorbers 
might be affected  by  different processes such as 
star formation, energy feedback, etc., and by different
 mixing mechanisms like 
those produced by tidal forces and mergers. These processes
depend on the history of evolution of each galaxy which may
be also affected by its environment (see Ellison \& L\'opez 2001).
 Hence, it is not simple
to envisage a model (e.g., Mathlin et al. 2001) to explain DLA properties and to relate them
to the characteristics of the associated DLA galaxies,
even more, if the internal structure of the systems and environmental effects
ought to be taken into account.

DLAs observations provide clues on the chemical evolution of the observable
neutral H mass in the Universe. Since the metallicity
of the ISM is the result of the  effects of 
different physical processes such as stellar evolution, mergers,
interactions, collapses, etc., with some of them depending also on the 
cosmology, it is relevant for a galaxy formation model to be able to
reproduce them. In this way, the individual and global
chemical   properties of the matter
can be used as tests for  galaxy formation and cosmological models.
Hydrodynamical cosmological simulations are adequate tools to tackle these
problems since the dynamical range that can be resolved
 allows the statistical description
of internal properties of the galactic objects such as mass distributions and
star formation histories, as well as mergers and interactions
(see Somerville et al. 2001 for a semianalitical approach).

Tissera et al. (2001, Paper I)
 used a chemical model coupled to a hydrodynamical
cosmological code  to analyse the 
metallicity properties of the interstellar medium (ISM) of the
galactic objects when mapped by random LOSs.
 These authors found
that the chemical abundances   obtained  in such a  way were comparable
to those measured from DLAs observations. Their results support the
hypothesis that at least part of the observed 
DLAs could originate in the building blocks of current
normal galaxies, in agreement with those found by Haehnelt et al. (1998)
from
a kinematic analysis.
In this paper we extend their work, making a more detailed analysis and
comparison between the simulated and new  observed DLAs.
We also analyse comparatively the properties of the stellar populations
intercepted by the LOSs and consequently associated to the
DLAs, and those of
the total stellar populations of  their host galaxies.


\section{Numerical Models}
The hydrodynamical chemical simulations analysed follow the joint evolution of 
dark matter and baryons within a cosmological context (Tissera et al. 1997),
including star formation  and chemical evolution. The hydrodynamical
equations have been implemented in the AP3M gravitational code 
(Thomas \& Couchman 1992) by using the Smooth Particle Hydrodynamics (SPH)
technique.
Baryons are initially assumed to be in the form of gas.
 Dense and gaseous regions in  convergent flows  are gradually
transformed into stars at different  star formation (SF)
episodes,
 according to the Schmidt law.
A star formation efficiency ($c$) has to be assumed to regulate this
process. 

The chemical model used in this paper
has been discussed in detail by Mosconi et al. (2001).
Briefly, it is a self-consistent implementation that  considers the chemical
evolution of the SP and 
ISM taking into account the contributions of 
different SP generations.
Type Ia (SNIa) and type II (SNII) supernovae  
are taken into account according
to stellar evolution models and chemical  enrichment yields.
We adopted the yields given by Woosley \& Weaver (1995) for SNII
and by Thielemann, Nomoto \& Hashimoto (1993) for SNIa.
We assumed a fixed Salpeter Initial Mass
Function with lower and upper mass cut-offs of $0.1 \ {\rm M_{\sun}}$
and $120\ {\rm  M_{\sun}}$, respectively.
Chemical elements generated in a given particle
are distributed  within
its  neighboring area, weighting each contribution with a
kernel function that  
depends on the relative distance
between gaseous particles.
 The distribution of metals
by using the SPH technique results in an effective mixing
mechanism that allows the enrichment of  regions that 
are nearby to star-forming particles (see Lia et al. 2002 for a different
implementation of the mixing process).

Although we stress the relevance of the energy injection into the
interstellar medium due to supernovae explosions
 in the formation of the galaxies, we have not included this
mechanism in this work.
Several attempts to implement energy feedback in 
SPH codes can be found in the scientific literature
(e.g., Katz 1992; Navarro \& White 1994;
Metzler \& Evrard 1994; Navarro \& Steinmetz 2000; Springel
2000), although they are still quite controversial.

The simulations studied in this paper followed the cosmological evolution 
 of  typical $5 \ h^{-1}$ Mpc  ($h=0.5$) cubic volumes  
represented by $64^3$ equal mass particles 
($M_{\rm p} = 1.4 \times 10^8 \ h^{-1} M_{\odot}$).
Initial conditions are consistent
with a Standard Cold Dark Matter (SCDM) Universe
($\Omega=1$, $\Omega_b=0.10$,  $H_0=100 \ h^{-1}\ {\rm km \ s^{-1}Mpc^{-1}}$) with cluster abundance normalization,
 $\sigma_8=0.67$ and $\Lambda =0$. We used a gravitational softening $\epsilon_{\rm g}=1.5$ kpc and 
a minimal smoothing length of $\epsilon_{\rm g}/2$
(see Section 3.1 for a discussion on numerical effects).

In Paper I, we analysed a set of three simulations 
(S.2, S.6 and S.7), each one representing 
different realizations
of the power spectrum run by Mosconi et al. (2001).
As shown by these authors, at $z=0$ these simulations reproduced
 galactic systems with mean metal abundances in 
agreement with observations.  
However, when the  abundance patterns of the stellar populations
in  the individual galactic systems
were analysed in detail, we found that none of them resembled fairly well
 that of
the Milky-Way, although averaged abundances are within observed values.
  Therefore, we run a new simulation (S.8)
with different SN parameters
in order to be able to have systems with  chemical abundance patterns 
similar to those of the Galaxy (i.e., we have increased $c$
and  the ratio between SNII and SNIa rates). For SNIa, we have adopted
a life-time for progenitor  binary systems
 of 0.5 Gyr. 
However, note that we do not know how common the Milky-Way abundance patterns
are among similar morphological type galaxies (Wolfe et al. 2000), 
so we consider that both
set of simulated objects can provide  useful information about the chemical
evolution of the structure in hierarchical clustering models. 
In this paper we will focuse the analysis on simulation S.8.

\section{Simulated DLA galaxies and DLA Systems}
 
Following the hypothesis that the progenitors of today normal galaxies
in hierarchical scenarios could be the host galaxies
of the observed DLAs, we
carried out a comparative analysis of 
the  mean chemical properties
 of the simulated DLA galaxies and 
DLA systems. The former are given by the ISM
and SP residing in dark matter
structures that are detected by an overdensity criterium. 
On the other hand, 
random LOSs are drawn through these structures
following the  observational procedure used to detect them.
Simulated DLAs correspond to those LOSs that map regions where
the neutral hydrogen column density $N({\rm H_{I}})$  is 
greater than $10^{20} {\rm atoms/cm^{2}}$.

\subsection{Galaxy-like Objects}

We identified galactic objects 
at different stages of evolution of the simulated volumes.
Firstly, a friend-of-friend method is used to isolate the higher
density peaks
in the mass  distribution, and then,  a density-contrast algorithm is
applied to find  over-densities with $\delta \rho /\rho \approx 200$.
The selected galactic objects are the building blocks of today
normal galaxies which grew  by accretion and/or mergers. The
rate of mergers 
as well as the distribution of merger parameters
 are  determined by the cosmological
model adopted.
 The main baryonic 
clump within a virialized structure will be, hereafter,
 referred to as the galaxy-like object (GLO),
and it constitutes what we also call simulated DLA galaxy.
GLOs have been truncated
at two optical radius, $2R_{\rm opt}$, where $R_{\rm opt}$ has been
defined as 
  the radius which 
encloses $83\%$ of the baryonic mass. 

For the purpose of  diminishing numerical resolution problems,
we analyse  galactic objects  with more than 200 baryonic
particles within their virial radius and in the redshift 
range $0.25 < z < 2.35$.
The final sample is made of 166 GLOs 
that satisfy the above conditions 
and have virial velocities
within
$\approx 100-250 \ {\rm km \ s^{-1}}$.
Owing to the fact that all particles have the same mass, the selected
objects have well resolved dark matter haloes (more than 2000 particles)
  providing  well-defined potential wells onto which, on its
turn,  
 baryons can settle  on (Steinmetz \& White 1997). Hence,
although the baryonic mass resolution is low 
compared to other simulations (e.g., Cen et al. 2003), 
the gaseous  density profiles are well described.
This is an important point since the star formation rate and,
consequently,
the metal production, is 
proportional to the gas density (e.g., Tissera 2000), and hence
it depends on how well the  profiles are described.

As a combined result of dynamical evolution, mergers and interactions,
the SF rate history of each galactic  object in hierarchical clustering
scenarios
can be described as a contribution of an ambient SF rate and a
series of starbursts (e.g., Tissera 2000).
In our models, the timing between starbursts are  given naturally
by the evolution of the objects in the hierarchical clustering
scenario adopted. 
As discussed by Tissera (2000 and references therein),
Tissera et al. (2001) and suggested by recent observations
(e.g., Barton, Geller \& Kenyon 2000; Le Fevre et al. 2000; Lambas et al. 2003) mergers and
interactions  play
a crucial role in triggering SF and consequently, may leave
important imprints in the chemical patterns of galaxies 
(Prantzos \& Boissier 2000). 
Our model provides a consistent description of these
processes as well as including the contributions
of SNIa and SNII consistently with the SF histories  of the objects. 

In order to understand
how metals are distributed within the GLOs,
we have considered 
concentric shells centred at the GLO
mass centre. The shells have an outer fixed radius
of $2R_{\rm opt}$ and an inner radius $r$, which
varies from $r=0$ to $r=10$ kpc.
For each shell, we define the ${\rm H_{I}}$ mass-weighted mean
abundance ratio of
elements K and J,
[K/J]$^{\rm gas}$  
as:

\begin{equation}
[{\rm K/J}]^{\rm gas}={\rm log} \frac{\sum_{i=1}^{ n_{\rm p}} {\rm K}_i \,  M^{\rm gas}_i}{\sum_{i=1}^{ n_{\rm p}} {\rm J}_i \,  M^{\rm gas}_i} - {\rm log}({\rm K/J})_{\odot}
\end{equation}

\noindent
where  $ n_{\rm p}$ is the total  number of gas 
particles belonging
to a GLO ,
$M^{\rm gas}_i$ the hydrogen remnant
in  the $i^{th}$
particle,  ${\rm K}_i$ and ${\rm J}_i$ their chemical abundances,
and $({\rm K/J})_{\odot}$ the corresponding solar abundance ratio.
Similar estimations can be defined for stars by summing up  
over the stellar populations.
These analysis provides us with information on the metallicity distribution
of the 
gaseous
and stellar components in each galactic system as a function of $z$.
It has to be noted that in these models
we are not able to distinguish between ionized and neutral gas, therefore 
 we assume that  both phases have the same chemical abundances.

\subsection{Simulated DLAs}

Observations of DLAs provide information on the 
chemical properties  of the ${\rm H_{I}}$ components belonging to 
structures lying along LOSs towards QSOs.
Specifically, DLAs are defined as those ${\rm H_{I}}$ components  with column density  
$N({\rm H_{I}})
higher than 2 \times 10^{20} {\rm atoms\ cm^{-2}}$ (Wolfe et al. 1986).
In the simulations, 
the ${\rm H_{I}}$ components of   the main baryonic clump (i.e., the GLO)
within the virial radius of the galactic object will be considered as  
possible absorbers.
Following Tissera et al. (2001), we use Monte Carlo technique to draw 
LOSs  through GLOs 
 and estimate the chemical properties of the ${\rm H_{I}}$ components with
 ${N({\rm H_{I}})> 2\times 
10^{20} {\rm atoms \ cm^{-2}}}$  along them.
For each LOS drawn through a GLO at a certain $z$
we estimate the chemical  abundances of the 
${\rm H_{I}}$ component along it
by applying Eq.1., but where the sum is now defined over the particles 
which 
are situated along a given LOS. 
Three different DLA mock
catalogs have been produced 
by generating three different random observers and drawing
a LOS through each GLO from each of the three observers.
In this way, we generated a total of 380 DLAs  
distributed between $z=0.26$ and
$z=2.3$.

Owing to geometrical considerations, random LOSs tend to map  
with higher probability the properties
of the outer regions of the absorbing systems, 
missing  information coming
from the central regions (e.g., Jimenez et al. 1999; Somerville et
 al. 2001; 
Savaglio 2000). Note that impact of geometrical effects
depends on the absorber redshift as we will discuss later on. 
It still remains to be answered how representative of the properties of their hosting
structures these observations are
(e.g. Pettini 2003).
This is one of the aims of this paper since the simulations
provide information on both the DLA system and its host galaxy.


For this purpose, we  estimate the impact parameter $b$ in kpc as  
 a function of $z$.  
The $b$ parameters found in the simulations take values
  from few up to $\approx 60$
kpc
with a mean  of $\approx 15 $ kpc,
again showing  the fact that 
 random LOSs tend to map preferentially
the outer regions of the baryonic main clumps. 
We found that for the absorbers at $z<1$, the percentage  with
$b \leq 5$ kpc  is $39\%$ of the total, while
at $z >1$ this percentage goes down  to $21 \%$.
In terms of the virial radius, we found that regions causing the 
absorptions are between  $5 \%$ to $25 \%$
 of the virial radius of the galactic objects.
This result is in marginal agreement with that reported by 
Haehnelt et al. (2000),
but it is roughly consistent with the expected scale-length of a centrifugally
supported disk if the specific angular momentum  of baryons is
assumed to be conserved during  disk formation.
This difference found between the results of Haehnelt et al. 
and ours may be due to the fact that our simulations include star formation.
The transformation of gas into stars has non-negligible effects on the gas 
dynamics, specially during violent events as reported by Mihos \& Hernquist
(1996) and Dom\'{\i}niguez-Tenreiro, Tissera \& S\'aiz (1998), among others.
 We also detect  a  slight trend for this
average percentage
to increase with $z$.

\section{Metallicity Evolution}

In this Section we analyse different statistical moments of 
the metallicity  
of the ISM  and the associated 
 SPs 
 intercepted by  random LOS.
We will analyse the variation of unweighted  
and mass-weighted  means  of 
different abundances with redshift. We will also apply linear regression
through the data to quantify possible  evolutional signal. 
The mass weighted 
mean
 abundances as a function of redshift
provide clues
on the global metallicity content of the Universe, while linear
regressions through the data or unweighted means give idea on the
changes in the chemical abundances of the individual galactic objects. 

The chemical elements
 more commonly used to assess the metallicity properties of
 DLAs are zinc (Zn) and iron (Fe).
The Zn is usually considered
a reliable tracer of metallicity since it is weakly depleted onto dust.
However, its poorly understood nucleosynthesis and the difficulty to
be detected at low metallicity make this element troublesome to interpret.
On the other hand, the Fe is heavily depleted onto dust grains but it is 
less biased than Zn against low metallicity DLAs, and its nucleosynthesis
is well understood.

The so-called $\alpha$-elements (e.g., silicon (Si), sulphur (S), calcium (Ca), oxygen (O))
are also observed in DLAs although some of them could be importantly
affected by dust depletion.
The  dust-to-gas ratio is
larger for refractory elements and varies with environment making
corrections complicated (Vladilo 2002).
The simulated sample is not affected by dust depletion or
obscuration, neither it is biased
against low or high metallicity column densities,
hence these simulations are a powerful tool
to explore the nature of DLAs.

In order to properly confront
 our results  with  observations,
we have gathered the
available observational data of DLAs
within our  redshift range of interest
and applied to them
 the same analysis performed
to the simulated DLAs.
We have chosen to use the data provided by Vladilo (2002) since they 
 include self-consistent
dust corrected values making them more suitable to constrain
the results obtained from our simulations which do not include dust effects.
 This observational sample includes data
from several authors and, in particular, overlaps with the data base
analysed by Prochaska \& Wolfe (2002).


\subsection{The Interstellar Medium}

For the purpose of assessing possible 
 signals 
of global or intrinsic evolution through the comparison of 
 means at different redshift intervals
(e.g., Prochaska \& Wolfe 2000),
 we split both the observed and simulated samples 
 into two redshift bins: $z_{\rm low}=
[0.25, 1.5]$ and  $z_{\rm inter}= (1.5,2.35]$.
Following Tissera et al. (2001), 
we first estimate at each analysed $z$
the unweighted mean 
$\langle {\rm [X/H]} \rangle =1/n\sum_n{\rm [X/H]}$ for  Zn,  Fe and Si abundance ratios
(represented
by X in the previous equation)
 for the   ${\rm H_{I}}$ along the LOSs intercepting the GLOs
(where $n$ is the number of LOSs at a given $z$).
The $\langle {\rm [Zn/H]} \rangle$ and $\langle {\rm [Fe/H]} \rangle$ values at each $z$ are
shown in Fig.1 as filled circles with error bars indicating 1$\sigma$
standard deviations. 
Regarding the dispersions in the simulated DLAs sample, we found
that they are comparable to the dust corrected data from
Vladilo (2002) as it can be also
appreciated from Fig.1.
Table 1 summarizes the observed estimations for the data presented by
 Vladilo (2002) 
($\sigma_{\rm z_{low}}^{\rm obs}$, $\sigma_{\rm z_{inter}}^{\rm obs}$) 
 and those corresponding to
the simulated DLAs 
($\sigma_{\rm z_{low}}^{\rm sim}$, $\sigma_{\rm z_{inter}}^{\rm sim}$)

We will refer to the  difference between 
the unweighted means of a given abundance ratio ([X/Y]) 
estimated  for the simulated DLAs at the two adopted  redshift intervals
as 
$\Delta {\langle {\rm [X/Y]}^{\rm sim} \rangle}_{\rm u}$.
In the case of [Zn/H], 
the difference $\Delta {\langle {\rm [Zn/H]}^{\rm sim} \rangle}_{\rm u}$  
 yields an intrinsic evolution 
of $-0.23 \pm 0.09$ dex for the simulated DLAs. 
A linear regression through the simulated sample
gives an  anti-correlation signal  with  
 a slope of 
 $d{\rm log[Zn/H]}/dz= - 0.26 \pm 0.04$ dex. 
In both cases, errors have been estimated by using the re-sampling
bootstrap technique with 500 random samples. 
Estimates of the
unweighted mean ${\langle {\rm [Fe/H]}^{\rm sim} \rangle}_{\rm u}$ at low and intermediate 
redshift bins 
and a linear regression show 
an intrinsic  evolution in the simulated sample of 
$\Delta {\langle {\rm [Fe/H]}^{\rm sim} \rangle}_{\rm u}=  -0.33 \pm 0.11$ dex,
and  $d{\rm log [Fe/H]}/dz=-0.36 \pm 0.05$ dex, respectively.
The linear regressions trough the [Zn/H] and [Fe/H] abundance ratios
for the whole sample of simulated DLAs are depicted in Fig.1.

Table 1 summarizes the linear regressions obtained for the simulated and
observed DLAs with and without dust corrections.
Table 2 gives  the differences between the unweighted means estimated at the two adopted  redshift bins
for different abundance ratios obtained from the simulated DLAs 
($\Delta {\langle {\rm [X/Y]}^{\rm sim} \rangle}_{\rm u}$).
This table also shows the corresponding values for the observed DLAs reported by Vladilo (2002)
with and 
without  dust corrections ($\Delta {\langle {\rm [X/Y]}^{\rm obs}_{\rm dust} \rangle}_{\rm u}$ and $\Delta {\langle {\rm [X/Y]}^{\rm obs} \rangle}_{\rm u}$,
respectively).

The cosmic mean metal evolution can be estimated by calculating 
the difference 
of  
${\rm H_I}$ mass weighted averages 
at the low and intermediate
redshift bins
($\Delta {\langle {\rm [X/H]}^{\rm sim} \rangle}_{\rm w}$).
 These estimations  yield an evolution of $0.51 \pm 0.39$ dex, 
 $0.85 \pm 0.48$ dex and $0.72 \pm 0.59$ dex for the 
[Zn/H],  [Fe/H] and [Si/H], repectively (bootstrap errors given, BE).
 Note that although the values are large, the BE errors  indicate that
they are statistically consistent with non-evolution within $2\sigma_{\rm BE}$.
Table 3 summarizes these differences for both, simulations
and observations.

Overall, we acknowledge higher evolution signal in our simulated DLA sample
than that obtained from the dust-corrected sample of Vladilo (2002),
although the bootstrap errors of the observations are quite high suggesting
that the sample suffers of low statistical number.
We will come back to this point in Section 5.

\subsection{Stellar Populations}

DLAs are ${\rm H_{I}}$ clouds 
detected in absorption providing no direct information on the SPs 
associated to the absorber. 
One of the advantages of our models  is that 
 we can compute
the metallicity properties of the stars situated along  
the  LOSs. 

In this case, we  find 
an intrinsic evolution signal of $0.23 \pm 0.13$ dex (BE)
for the [Zn/H], 
 $0.16 \pm 0.13$ dex for the [Fe/H] and $0.15 \pm 0.17$ dex for the [Si/H],
 over the $z$-range of interest.
In Fig.2, we plot the 
unweighted
mean abundance ratios at each analysed redshift (filled circles)
and the dust-corrected data of   Vladilo (2002; open pentagons).

We also define the stellar mass weighted 
mean
metallicity of 
the SPs along LOSs.
The
difference between these
 mean values at the low and intermediate redshift bins  
indicates  a global  evolution of $1.00 \pm 0.48$ dex (BE),  $0.84 \pm 0.34$ dex and 
$0.92 \pm 0.57$ dex for
the [Zn/H], [Fe/H] and [Si/H], respectively.

The degree of intrinsic and global evolution detected in the simulated SPs and the ISMs
are comparable (within $2 \sigma_{\rm BE}$) suggesting that both mass components
reflect similarly the rate of enrichment of the galactic objects
as they evolve.
In order to quantify if they
have the same   amount of enrichment, we calculated the difference
between the SP abundances and those corresponding to  the gaseous component 
along each simulated  LOS. 
For the unweighted means, we found that stars are
always less metal-rich  by, on average,  $\approx 0.50$ dex  ($\approx 0.40$ dex for [Zn/H] and $\approx
 0.70 $ dex for [Fe/H]) than the ISM where they inhabit at any  redshift,
while the weighted means are very similar. 
These results indicate
that, in these models, there are stellar populations that have lower metallicities compared to those of their  associated DLAs,
 but
represent a small percentage of the total stellar mass;
 while a larger
fraction  of stars is
as metal rich as the ISMs they are embedded in,  at both redshift intervals.
It is very interesting to note that the importance of this segregation in the abundances of the SP and the ISM
identified along random LOSs depends on the chemical element considered. The so-called
$\alpha$-elements show a smaller difference than those mainly produced by SNIa.
The difference is originated in   the distinct star formation histories of the regions and
the time delay between the two kind of SN events (Tissera et al. 2002).

Finally, we find that  the dispersions of abundances of the SPs are higher than those
measured for the ISMs (see Table 1) along LOSs, showing that abundances of  stars in regions mapped by those LOSs
are more heterogeneous (for example, the
[Fe/H] show  $\sigma_{\rm z_{inter}}=0.74$ 
and $\sigma_{\rm z_{low}}=0.85$, and for  [Si/H] we have  $\sigma_{\rm z_{inter}}=0.66$ 
and $\sigma_{\rm z_{low}}=0.70$).

The simulated abundances and their dispersions are a consequence of the variety
of GLOs with different evolutionary histories. The
fact that their ISMs when mapped by random LOSs
 show similar chemical properties to those  measured in DLAs 
supports the hypothesis that DLAs absorptions can be actually produced by 
  a mixture
of galaxy types. Since we are
simulating typical field regions of the Universe, the majority of the GLOs
are supposed to represent typical spirals at $z=0$ with some contributions from
other morphological types (Dressler 1999). 
As we go to  higher $z$, the substructures
that have merged to form them start to emerge as separate entities. 
The mixture
of the properties of these progenitor objects gives rise to the
chemical characteristics of  the
simulated absorption clouds and the SPs.


\section{Dependence of  metallicity  on  column density}

There is evidence showing that high metallicity
and high column density DLAs are missing from the observational data.
However, it is not clear if this is owing to   
the depletion of metals  onto dust from the
diffuse phase, to geometrical effects, or if they
are not present at all in nature.
Boiss\'e et al. (1998) estimated an observational region where all observed
DLAs lay, defined as $F={\rm [Zn/H]} + {\rm log} N({\rm H_I})$ with $18.8 < F< 21$.
In order to mimic in the models 
the observational lack of high metallicity and
high column density DLAs,  we followed 
Prantzos \& Boissier (2000) and  applied this observational  filter to the simulated sample.
We found that $30 \%$ of the sample is cast out, with
$19 \% $ having $F >21$. These DLAs are high metallicity and
high column density absorbers.

Estimations of the evolution signals for the filtered simulated DLAs sample
indicate that both intrinsic and global evolution are much weaker than those obtained from
the original simulated sample, as it can be seen from Tables 1,  2 and 3.
The difference of unweighted means
for the filtered sample are 
$\Delta {\langle {\rm [Zn/H]}^{\rm sim}_{\rm fil} \rangle}_{\rm u} = -0.03 \pm 0.09$ dex,
$\Delta {\langle {\rm [Fe/H]}^{\rm sim}_{\rm fil} \rangle}_{\rm u} = -0.21 \pm  0.08$ dex and
$\Delta {\langle {\rm [Si/H]}^{\rm sim}_{\rm fil} \rangle}_{\rm u} = -0.19 \pm 0.08$ dex,
while for the mass weighted means are
$\Delta {\langle {\rm [Zn/H]}^{\rm sim}_{\rm fil} \rangle}_{\rm w} = -0.14 \pm 0.04$ dex, 
$\Delta {\langle {\rm [Fe/H]}^{\rm sim}_{\rm fil} \rangle}_{\rm w} = -0.24 \pm 0.21$ dex and 
$ \Delta {\langle {\rm [Si/H]}^{\rm sim}_{\rm fil} \rangle}_{\rm w} = -0.27 \pm 0.09$ dex.
We have also carried out estimations of evolution by applying only the upper limit of
the observational filter finding very similar results. 
Hence, the high column density and high metallicity DLAs are the ones imprinting 
signals of intrinsic and global evolution. 

The physical processes that cause  the non-detectability
of such  HI clouds remains to be explained. 
 Although dust can
account for it as it  has been reported by several authors (e.g., Hou et al. 2001),
recent results from a study of DLAs against radio-selected QSOs
(Ellison et al. 2001)  suggest that dust may not be 
as important as previously thought since high column density DLAs have not
been detected yet in this sample.
In the simulations we found  that  the percentage of DLAs with these
characteristics (i.e., outside the upper limit of the observed filter)
 is only $3\%$ and $ 16 \%$ at  $z >1 $ and $z < 1$,  respectively.
Hence, according to our results,  the radio-selected  sample is still small to discard geometrical effects as a possible cause,
since most observed DLAs are at $ z> 1$ (Cen et al. 2003). 
Moreover, because the  filtered simulated DLAs are more consistent with the dust-corrected
sample, our findings do not support the hypothesis that this
upper observational limit is caused by dust obscuration but they
favour    
 geometrical effects
\footnote {An alternative option is proposed by Schaye (2001b) who claims
that the maximum ${\rm H_{I}}$ column density is determined by the conversion to
molecular hydrogen. The fraction of molecular H correlates with
 dust content and total hydrogen, and anticorrelates with the intensity
of the incident UV radiation. However, this author
 does not discard  dust effects
as the possible explanation of the apparent lack of systems with low column
density and high metallicity which cannot be accounted for by his models. }.

We can use the simulations to explore the characteristics of  
DLAs cast out by the filter.
In Fig.3, we plot the stellar mass along the LOSs
versus the filter $F$. The dashed-dotted lines depict the limits
established by the observational filter. Stellar populations
associated to DLAs have been represented
by filled triangles. Sub-DLAs ($10^{19} <  N({\rm H_I})  < 2 \times 10^{20}
\, {\rm atoms \ cm^{-2}}$) have been also included in this figure but
they will be discussed in Section 7.
As it  can be seen from the distribution of the filled symbols, 
there is a clear correlation between the stellar
mass associated with a given DLA and the filter characteristics of that DLA.
There are very few DLAs with $F <18.8$,
and  all simulated DLAs with $M_{\rm stars} >10^8 \ {\rm M_{\odot}}$ have
$F >21$.
Therefore, in our models, this observational filter directly translates
into a stellar mass cut-off. 
As it has been shown, most LOSs map the external regions of DLA galaxies.
However those with $F >21$, which are  comparable to  those missing from observations,
have impact parameters 
 $b < 5$ kpc.  Hence,
 LOSs with $M_{\rm stars} >10^8 \ {\rm M_{\odot}}$   intercept
the central regions. Although the percentage of such 
LOSs is low ($\approx 7\%$ of the total), 
it is  enough to imprint  global evolution
with redshift in our simulations. It has  been already mentioned 
that 
 simulated DLAs with low impact parameters  are more common at $z <1$ because of 
geometrical effects.
These results suggest that some effort should be devoted to enlarge   
 the sample of observed DLAs at low redshift since it could be profitable in order
to test the existence of DLAs with high $N({\rm H_I})$ and high metallicity.
Yet, another possibility  could be  that the gas densities 
might  have been
lowered down by  SN energy injection (not treated in this work), and consequently
DLAs with these features are not present at all in
nature as suggested by Ellison et al. (2001).

\section{The $\alpha$-elements }

The abundance ratios of the so-called $\alpha$-elements (i.e., Si,
S, C, O) in DLAs have been subject of many studies (e.g., Lu et
al. 1996; Pettini et
al. 1997; Pettini et al. 2000) because
they are  thought to be
 good indicators of the  SF history of DLA galaxies. 
This argument is based on the fact that the abundance
patterns of the $\alpha$-elements in the Milky-Way
show enhancement for metal poor stars ([$\alpha /$Fe]$ = + 0.5$ dex at 
[Fe/H]$ \leq -0.66$) and then, a decline towards solar values for higher metallicities.
This change of behaviour can be explained by taking into account  the
time-delayed contribution of SNIa to the chemical enrichment of the ISM.
This type of  SNe are the main producer of Fe which is ejected into the ISM after
a time-delay  equal to the life-time of the binary systems
from which they arise
(Thielemann et al. 1993).
The timing of these chemical contributions are determined by the
 particular history
of star formation of each galaxy.

It has been found that the presence of  dust in DLAs
 affects the determinations of 
 $[\alpha/$Fe] ratios in a complicated way. Estimations of dust-corrected
abundances yield  nearly solar values (Vladilo 1998; Vladilo 2002).
Recently, Ellison \& L\'opez (2001) found a pair of DLAs selected via
radio observations which show
solar and  sub-solar [$\alpha/$Fe] abundances.
Sub-solar values would suggest that the ${\rm H_{I}}$ clouds have been 
mainly chemically enriched by SNIa.
Based on the discussion presented by Vladilo (1998) and Hou et al. (2001),
among others, we focus on the comparison of our results with dust-corrected observations
since metal depletion  onto dust has not been considered in our simulations.

First observational results of $[\alpha/$Fe] in DLAs shown
  some enhancements and a 
lack of trend with increasing redshift and metallicity (Pettini et
  al. 1997).
However, recently, a behaviour consistent with a general trend for 
decreasing $\alpha$-enhancement with metallicity has been reported 
by Centuri\'on et al. (2000), Molaro et al. (2001) and Vladilo (2002). 
In particular, Molaro et al. (2001) estimated a change of $-0.36$ dex from 
a regression analysis of [S,O/Zn] providing the first results
that give hints on metallicity evolution in the $\alpha$-elements.
In a DLA sample with redshifts $> 1.5$, Prochaska \& Wolfe (2002) found that
dust can affect strongly the measurements in those absorbers with
[Si/H]$>-1.5$.
They found that DLAs at low metallicity, which can be considered dust-free,
have significant $\alpha$-enrichment, as
indicated by the [Si/Fe] values which exhibit a plateau of $\approx 0.3$ dex at
[Si/H]$< -1.5$ dex. The increase in [Si/Fe] at higher values of [Si/H] can be
explained by dust depletion. Thus, when allowing for this effect, these
authors found that the abundances of different systems are quite homogeneous.
 
In Fig.4(a)
we show the distribution of [Si/Fe] versus [Fe/H] 
for the simulated DLAs, the dust-corrected data
of Vladilo (2002).
We can see that the abundance ratios 
for the simulated
DLAs take values  
from solar  up to [Si/Fe]$\approx 0.30$ dex, in very good agreement with
the observed dust-corrected abundances.  Note also
that there are some simulated DLAs with low
 metallicities and (sub)solar [Si/Fe],
indicating a main  enrichment by SNIa.
This behaviour is consistent with that reported by some observations (Ellison
\& L\'opez 2001)
which show sub-solar $\alpha$ values for some low metallicity absorbers,
and has arisen from the natural evolution of the GLOs which takes
into account infall of material from the dark matter halo as well as inflows
within a given GLO.
In order to assess the presence of evolution of the $\alpha$-elements
with metallicity, we performed a linear regression through the
simulated DLA data finding $-0.15 \pm 0.01$ dex
from [Si/Fe] values.
These estimations have been performed with those
simulated DLAs that satisfy the  usual  definition  
of $N({\rm H_{I}}) > 2 \times 10^{20}\
{\rm atom/cm^2}$.

In Fig.4(b), we have plotted the DLAs that are within
the observational filter defined in the previous Section.
Note how the filter tends to cast out the simulated DLA
at the plateau of the relation, resulting  in a better agreement with observations:
 $d{\rm log [Si/Fe]/d[Fe/H]=-0.23 \pm 0.01}$ dex.
This effect is caused because the observation filter casts out
the high metallicity and high column density ${\rm H_{I}}$ clouds but
by using the [Zn/H] abundance as an indicator of metallicity.
This element is produced by SNII and has a different ejection timescale
than Fe which is mainly produced by SNIa. This different timing
between the two ejecta types can produce  clouds with
different relative  enrichments (Tissera et al. 2002).

Finally, Fig.4(c) shows the unweighted mean $\langle {\rm [Si/Fe]} \rangle$
of the simulated 
DLAs at each analysed $z$ as a function of the metallicity given
by the corresponding unweighted mean $\langle {\rm [Fe/H]} \rangle$,
where the
error bars represent the standard dispersions.
As it can be appreciated from this figure,  the dispersions for the 
$\langle {\rm [Fe/H]} \rangle $
 at each redshift are very large and more important than those of 
the $\langle {\rm [Si/Fe]} \rangle$ (Prochaska \& Wolfe 2002).
Hence, while absorbers have more homogeneous ${\rm [\alpha/Fe]}$ 
abundances
at a given redshift, they
show 
very different Fe content. This is the result of their different
histories of evolution and star formation which determined
the timing between SNIa and SNII contributions to the ISM.

Another way of assessing the differential enrichment by SNII and SNIa is
by analyzing the evolution of [Si/Fe] as a function of redshift (Fig.5).
As it can be seen, the agreement with the dust-corrected data from Vladilo (2002)
is very good, in the amount of enrichment, dispersion and trend.
A linear regression through the simulated data yields
$d{\rm [Si/Fe]}/dz= 0.05 \pm 0.01$ dex ($0.04 \pm 0.01$ dex for the filtered sample).
The possitive correlation is due to the increase of Fe content 
as the stellar populations get older and start to produce
SNIa.

Finally, in Fig.6 we plot the [Si/Fe] ratio as a function of
the metallicity [Fe/H] for the ISM and SP of DLAs
and Sub-DLAs; the discussion of the later ones 
is postponed till the next Section.  
Regarding the $\alpha$-enhancement of the SPs associated to
DLAs (filled stars), we can see that 
their [Si/Fe] abundances are higher than the corresponding ISM
and most of them formed in the higher
$\alpha$-content levels, implying that the regions mapped by the random LOSs 
are those that have experienced a burst of SF after which the
SF activity has proceeded in a more quiet way (see Tissera et al. (2002) for 
an analysis of the age of the stellar population in DLA systems).
This fact is also the cause of  having a SP less rich in Fe than
their ISM along the LOSs, since as the gas is enriched by SNIa from the old SP,  
on average, 
no important star formation locks  this material into stars.
The $\alpha$-content of the SPs does not show any trend contrary to 
their ISMs as mapped by DLAs. 

\subsection{Brief discussion on the meaning of the statistical analysis}

In this paper we have applied the two commoly used statistics to
assess evolution in the metallicity content of ${\rm H_{I}}$ clouds:
differences of means at different redshifts and linear
regressions.
These two statistics have been applied to the simulated DLAs and
the observational data presented by Vladilo (2002).
A close inspection of these results presented in Tables 1, 2 and 3
reveals interesting features that we would like to discuss.
We have chosen to show the bootstrap errors which give
an indication of the statistical significance of the relations,
but, the standard dispersions  
in simulated and observed DLAs are high (Table 1),  making difficult to
extract clear results.

Let us first look at the observations.
This analysis is focused on the [Fe/H] ratio, since iron is
much more affected by dust depletion than other chemical elements.
The differences $\Delta {\langle {\rm [Fe/H]}^{\rm obs}\rangle}_{\rm u}$
and $\Delta {\langle {\rm [Fe/H]}^{\rm obs}\rangle}_{\rm w}$ for the observed DLA sample
without dust correction are pretty similar, showing negligible evolution
both for unweighted and mass weighted means.
When dust corrections are incorporated, the evolution shown by the unweighted means
remains almost unchanged, while the difference between mass weighted means
increases to $\Delta {\langle {\rm [Fe/H]}^{\rm obs}_{\rm dust}\rangle}_{\rm w} = -0.30 \pm 0.35$ dex.
However, this value cannot be considered as indicative
of the presence of some evolution because of its large error.
This lack  of evolution is consistent 
with the negligible signal found by Prochaska et al. (2002) when
applying these two different statistics
to a DLA sample that covers a redshift interval twice as large than
the one considered in the present work. 

However, when considering the slope obtained from 
the linear regression through the observations, we found a
more statisical significant signal for intrinsic evolution,
 $d{\rm log [Fe/H]}/dz = -0.17 \pm 0.11$ dex, which increases 
 to $-0.31 \pm 0.14$ dex, when dust corrections
are applied. These values are pretty similar to those obtained by Vladilo (2002)
from the analysis of these data on a larger redshift range.
Hence, we obtained different results for the metal evolution in the 
observed sample, specially for 
the intrisic evolution, depending on  which statitics is  applied.

A similar behaviour is detected for the simulations, 
where more statistically significant signals are found when
linear regressions are used. 
 As it can be appreciated from Fig.1 and Fig.5, 
[Zn/H], [Fe/H] and [Si/Fe] ratios  for  
 both simulated and observed DLAs have
increasing dispersions with redshift, leading to a picture
that looks likely triangular.
In the simulations, this shape is caused by the combination of  enrichment
of the ISM of the building blocks and 
 the continuous gas infall which gives rise
to metallicity gradients. These gradients caused the  
 large abundance dispersions detected in the simulated DLAs, since
random LOSs  mapped different regions of the building blocks
(see Section 8).
The high metallicity and high column density simulated DLAs help to
increase continuously the upper enveloped abundances of the triangle, this is
why the filtered simulated sample shows less evolution when 
evaluated by differences of means. However, for the linear regressions
the DLAs cast out by the filter are too few to actually change the
general trend.
Finally, the fact that the  filtered sample reproduces 
the dust-corrected observed trends provides  
 evidences for the upper limit of the
observational filter not to be determined by dust absorptions.

\section{The impact of sub-DLAs}

Recent observational results of low density ${\rm H_I}$ clouds along 
LOSs, the so-called sub-DLAs,
 obtained by P\'eroux et al. (2001,
and references therein) show that clouds with $N({\rm H_{I}})$ 
in the range $10^{19} - 2 \times 10^{20} {\rm atom  \
cm^{-2}}$ contribute with an important fraction of the neutral gas of the
Universe,
specially at $z >3.5$.

Sub-DLAs   
 are associated with  regions of  lower stellar-mass
content than DLAs, as it is shown in Fig.3.
However, we find sub-DLAs to have a  negligible contribution to 
 the global metallicity evolution of the simulated box
 since in our models, such  evolution is  
mainly driven  by the high metallicity and high column density ${\rm H_{I}}$ clouds.
If sub-DLAs are incorporated to our calculations, we find that
they affect the estimation of the intrinsic evolution making the
signal even more stronger for both the gas and the stellar
population.
Hence, in the redshift range studied in this work, sub-DLAs seem  to
have no impact on the estimation of chemical evolution of the Universe but
they should be taken into account  to study the evolution
of the chemical properties of the ISM and SP in  DLA galaxies.

In fact, 
if  sub-DLAs are included in the  estimations of [Si/Fe] versus
[Fe/H] the metallicity evolution signal is even clearer, 
and an abundance pattern similar to that observed in the Milky-Way 
is identified in the distribution as it can be seen  in Fig.6.
 In the case of the filtered sub-DLA/DLA sample
we found metallicity evolution of $-0.29 \pm 0.03$ dex (BE). 
 The stellar populations associated
to sub-DLAs  tend to show  mild $\alpha$-enhancement and
the total filtered sample shows weak evolution with
metallicity: $0.07 \pm 0.03$ dex (BE).

The following step would be to disentangle if sub-DLAs are mainly
associated with small objects or  with the very outskirts
of  DLA galaxies.
Firstly, we probed for a dependence of the $N({\rm H_{I}})$ with
the virial velocity of the associated DLA galaxies finding that
DLAs and sub-DLAs are thrown out from a similar mixture of
galactic halos with no dependence on virial  velocity.
 Secondly, we looked at the relation between 
the impact parameter $b$ and 
$N({\rm H_{I}})$ (Fig.7).
 A clear trend is found for the higher impact
parameters to be mainly associated with sub-DLAs. Note that sub-DLAs also
trace the intermediate regions ($5 \ {\rm kpc}   < b < 8\ {\rm kpc}$).
Although this correlation is already well-known, we  have
estimated it in order to show that these model and simulation 
are able to reproduce them.

According to these results, both DLAs and sub-DLAs could be tracing
the ISMs of the same DLA galaxies with sub-DLAs providing more information
on the outskirts of the galactic objects. However we note that, in our models,
 we may be 
overestimating the gas column densities and consequently, underestimating
the impact parameters due to the well-known problem of excessive angular
momentum transfer from  baryons  to the dark matter
component. However, this effect works in the sense of erasing
the low density region contributions by making the objects more concentrated
with their gas density
artificially higher in the centre. 
Hence, the fact that we get a signal of this kind suggests that this
could be a real physical effect.
In that case, sub-DLAs should be included in the analysis of the chemical
evolution of the ISMs.

Finally,  stars associated with sub-DLAs 
show  $\alpha$-enhancement (Fig.6(b)), 
suggesting that the low-metallicity population in
the galaxies such as the Milky-Way, could have been formed in the outer
and intermediate regions of
protogalactic objects. Note that our results do not imply 
 any constrain on  the mass of such protogalaxies, actually we find 
no dependence at all on the virial mass.

Ionized gas may be significant in systems with low
$N({\rm H_I})$ which can be much more affected by ionization radiation (Viegas 1995).
Thus, the analysis of sub-DLAs should be taken
only as indicative since, in our model, the chemical abundances were obtained by 
assuming all gas in neutral phase.

\section{DLA Galaxies}

In this Section we focus on the analysis of host galaxies of the
simulated DLAs.
All the galactic systems analysed have been resolved numerically with more than
2000 particles within their virial radius and with a spatial resolution of
$1.5 $ kpc. 
 A detail description of these simulated galaxies
has been presented in Section 3.1.

Recent observations have  improved the statistics of
DLA galaxies (Rao \& Turnshek 2000; Nestor, Rao \& Turnshek 2001), although
it is still difficult to assert which sort of systems are responsible
for the absorption features in the spectra of QSOs at different redshifts.
 There are 
now approximately 10 DLA galaxies observed at $z <1.65$ (Le Brun et al. 1997;
Rao \& Turnshek 2000). These observations show systems with  
different morphologies and luminosities including  a significant
number of dwarf and low surface brightness galaxies, albeit also
spirals.
Turnshek, Rao  \& Nestor (2001) compared properties of these
DLA galaxies with those of local galaxies.
We will use this observations to confront with the properties of the GLOs
which host the simulated DLAs.

Our model allows to estimate the magnitudes of the GLOs in 
different wavelengths by using population synthesis model (Tissera et
al. 1997)
such as  GISSEL 98 (kindly provided by G. Bruzual). 
For each SF episode we estimate the flux distribution according
to its age and metallicity. Then, we 
sum up the contribution of all SPs belonging to a GLO at a certain
redshift in order to estimate its total luminosity.
In Fig.8,
 we plot the blue absolute magnitude of the GLOs at $z\le 1$,
distinguishing between those that are within (filled circles) and
outside (open circles) the observational filter.
Sub-DLAs have been included for completeness.
>From this figure we observe that those GLOs which give origin to
DLAs/sub-DLAs within the observational filter show a
 trend with magnitud similar to that found by 
 Turnshek et al. (2001)
for both DLAs and  local galaxies, in the sense that
the brighter the galaxies, the larger the impact parameter of the
associated DLA. 
The inclusion of GLOs cast out by the observational filter
 tend to erase the correlation  signal because the small $b$, which
are mainly related to DLAs with high metallicity and high ${\rm H_I}$ 
column density,
are associated
to galaxies of any luminosity; while larger impact parameters are
related to the brighter GLOs.

We also estimate the total blue luminosity of  GLOs at different $z$. 
As it can be appreciated from Fig.9, most of the building blocks producing
the absorptions have $(L/L^{\star})_{B} < 0.5$ 
as shown by the median at each $z$ 
(filled circles), in agreement with
observed DLA galaxies at $z <1$.
The dispersion found in the simulated ratios agrees well with
the observed fact that DLA galaxies cover a large range of luminosities.
We also found a trend for 
DLA galaxies to increase their tipical luminosity with redshift.
Note however that numerical resolution could affect this  last result.

The information on the chemical properties of the ISMs and the SPs
provided by the LOSs suggests that the simulated DLA galaxies at different
redshifts have metallicity gradients. 
In order to assess how fairly random
LOSs properly sample the chemical properties of the DLA galaxies,
we estimate the abundances of the ISMs and SPs in each GLO 
in concentric shells as explained in Section 3.1. We calculate averages
over the abundances within each shell in GLOs at each analysed $z$.

Fig.10 shows the mean values over GLOs  as a function of $z$ for the concentric shells
with  inner radius varying from $r=0$ kpc to $r=10$ 
kpc and the outer
one fixed at $r=2R_{\rm opt}$ (i.e., the estimations at $r=0$ implies
that all the mass associated to the simulated galaxies has been taken into account).
As it can be seen from this figure, the gas in the external regions is
less metal enriched than the material in the central areas confirming
the existence of metallicity gradients in the ISM of GLOs at all redshifts.
Linear regressions through the simulated data  yield 
evolution signal in the chemical content of the ISM of the whole GLOs
(i.e., $r > 0$ kpc) and of the ISM outside the very central regions (i.e., 
$r > 3$ kpc)  of $-0.32 \pm 0.06$ dex (BE) and $-0.23 \pm 0.03$ dex (BE)
 for  [Fe/H] (similar trends have been found for [Zn/H]).
This estimation agrees with that obtained from the linear regression of
the simulated DLAs.  For the ISM at $r >6$ kpc and $r >10$ kpc
 we get $-0.26 \pm 0.20$ dex (BE) and $-0.07 \pm 0.26$ dex (BE),
 implying that statistically, there is no evolution in the metal content
 of the intermediate
and external regions of the galactic structure, in agreement
with the filtered simulated DLAs. 
According to these  results we predict that
the observed global metallicity of the DLA galaxies might be underestimated,
since random LOSs tend to map
the external regions, 
 and
 no global evolution would be detected unless  
central areas are taken into account.

A similar analysis has been carried out for the SPs within each shell
finding that stars in the outer parts are less metal-rich than those
in the inner regions (Fig.11). 
Note that stars with $ r< 6$ kpc are the ones pulling up the averages
to nearly solar values. In this case, the gap between the metallicity of
stars in  the
inner regions and that of the rest of the GLOs is larger than those 
 of the ISM, 
where there is a continuous decrease of metallicity with radius.
Stars in the outer regions also show  
higher $\alpha$-abundances as expected.
We stress the fact that in the GLOs there are primordial stars
with very low Fe content ([Fe/H] $< -3$) and high $\alpha$-enhancements
but when averages are estimated, higher metallicity SPs are more
massive and dominate the estimations.
Finally, note that the mean abundances of the SP of the whole systems (i.e.,
inner radius $r=0$) are higher than those corresponding to the 
ISM. This is reflecting the fact that in hierarchical clustering scenarios
there is a continuous infall of pristine material from the  galactic halo
and that stars tend to form in the higher density regions where the metallicity
is higher.

\section{Conclusions}

In this work we studied the possibility that the progenitors
of current normal field galaxies are  DLA galaxies,
host of  absorbing ${\rm H_{I}}$ clouds in the context of hierarchical 
clustering, where galaxies form from the merger of smaller substructures.
DLAs observations provide information on the ISM of the host objects.
In our models, we  have 
 the comprehensive information of the ISM and SP of
the systems, allowing the performance of a  consistent study  in order
to obtain robust conclusions on the possible nature of DLA galaxies.

We constructed mock catalogs of DLAs by drawing random LOSs through
the galaxy-like objects at different $z$ in hierarchical clustering scenarios 
and compared the chemical
properties  of the simulated ISM when mapped by LOSs with observed DLAs. 
We then studied the properties of the simulated DLA galaxies and assessed 
at which extent random LOSs can properly extract information on both
the chemical properties of the galaxies as a function of $z$ and 
the global evolution of the Universe.

>From our simulations, we find

   \begin{itemize}

\item  the [Fe/H], [Zn/H] and [Si/H]  ratios of the simulated DLAs to match
the observational range determined by DLAs with similar dispersion.
In the case of the $\alpha$-elements the simulated abundances are 
consistent with the dust-corrected data;

\item intrinsic and global evolution for  [Fe/H], [Zn/H] and [Si/H] with 
$z$. However, when the observational filter suggested by Prantzos \& Boissier
(2000)
is applied, mild evolution consistent with observations is recovered;

\item  that  the  evolution with $z$ is mainly driven by the high
metallicity and high column density DLAs albeit these DLAs represent a small
percentage of the total number of LOSs ($< 10 \%$);

\item $\alpha$-element filtered abundances to evolve with metallicity  in 
agreement 
with recent observations (e.g., Molaro et al. 2001; Vladilo 2002);

\item evidences that the observational filter 
may not be reflecting dust absoption but other effects
such as geometrical ones;

\item a trend for [Si/Fe] to correlate with redshift as
it is expected by taking into account the different
ejecta timing of SNII and SNIa;

\item  sub-DLAs might be providing information on the chemical
properties of the external region of DLA galaxies, but they
do not affect the determination of global evolution in the metal
content of the simulated box, within $0.26 < z < 2.34$;

\item that the building blocks in hierarchical clustering scenarios can reproduce the 
magnitude-impact parameter correlation of observed DLAs for $z<1$;

\item no correlation between neither 
the impact parameters nor the  $N({\rm H_{I}})$
with virial velocity;

\item that metal-poor stars could be forming in the outer regions of galaxies
at any $z$ and the associated ISM might be tested by sub-DLA observations.

   \end{itemize}

These findings imply
 that the building blocks of current normal galaxies in  hierarchical
clustering models
could be host structures responsible of the DLA and sub-DLA systems and
that DLA galaxies  might  represent a variety of galaxy
morphologies with mean luminosities increasing with
respect to $L^*$ at $z=0$ with redshift.
Despite geometrical effects, 
a small percentage of 
high metallicity and high column density $N({\rm H_{I}})$ should 
be identified, unless supernova energy feedback works to lower the central gas
densities.
Since the  chances to intercept the central regions decrease
with redshift,  efforts should be put on detecting them
at $z<1$.
Metallicity gradients of the ISM of galaxies at different
redshifts may be estimated, on average, by looking at the range
of the metallicities set by DLAs and sub-DLAs. However
current observations would  be preferentially constraining the metallicity
distribution for $r > 5$ kpc.

The observed anti-correlation between impact parameter and magnitude
might   be the result of
the particular history of formation of each galaxy which host a DLAs or
sub-DLAs. However, the observational bias related to the lack of
high column density and high metallicity column density works in the sense
of supporting the anti-correlation by sweeping away the 
very low impact parameter contributors.

Stars in the simulated DLA galaxies (i.e., galaxy-like 
objects) are, on average,  more metal-rich at the centre and  metal-poor in 
the outskirts independently of $z$. 
Hence at $z=0$, a galaxy, which is the result of a merger sequence,  
can have old and young metal-rich
(and metal-poor) stars.
These findings have consequences for the understanding of the origin of the 
abundance pattern of the Milky-Way, since, according to our models, 
an important fraction of the low metallicity stars are situated
 in the outer ($r >6$ kpc)
regions at any redshift; stars with nearly solar values 
are present at any redshift, but in the central regions.
Hence, according to our results, 
the fact that observed DLAs yield low metallicity values
does not imply that the associated DLA galaxies cannot have
a metal rich population.
The abundance characteristics of DLAs and Sub-DLAs are 
the results of SNIa and SNII contributions plus gas infall.
The chemical characteristics of the SPs associated with simulated DLAs are consistent with having been formed in starbursts followed by quiescent periods
of SF. The fact that we always identify low metallicity stars 
in the outskirts and no important evolution of the metallicity in these
regions could be showing the action of continuous
infall.

Future works will be focused on the improvement of the chemical code
and
the performance of simulations with higher numerical resolution.


We thank the anonymous referee  for a carefull  reading of
the manuscript and
useful suggestions that helped to improve our results.
We thank the Max-Planck Institute for Astrophysics for 
the hospitality during PBT  visit where this manuscript
was finished.
This work was partially supported by the
 Consejo Nacional de Investigaciones Cient\'{\i}ficas y T\'ecnicas,
Agencia de Promoci\'on de Ciencia y Tecnolog\'{\i}a,  Fundaci\'on Antorchas
 and Secretaria de Ciencia y
T\'ecnica de la Universidad Nacional de C\'ordoba.



\clearpage



\begin{figure}
\includegraphics[width=84mm]{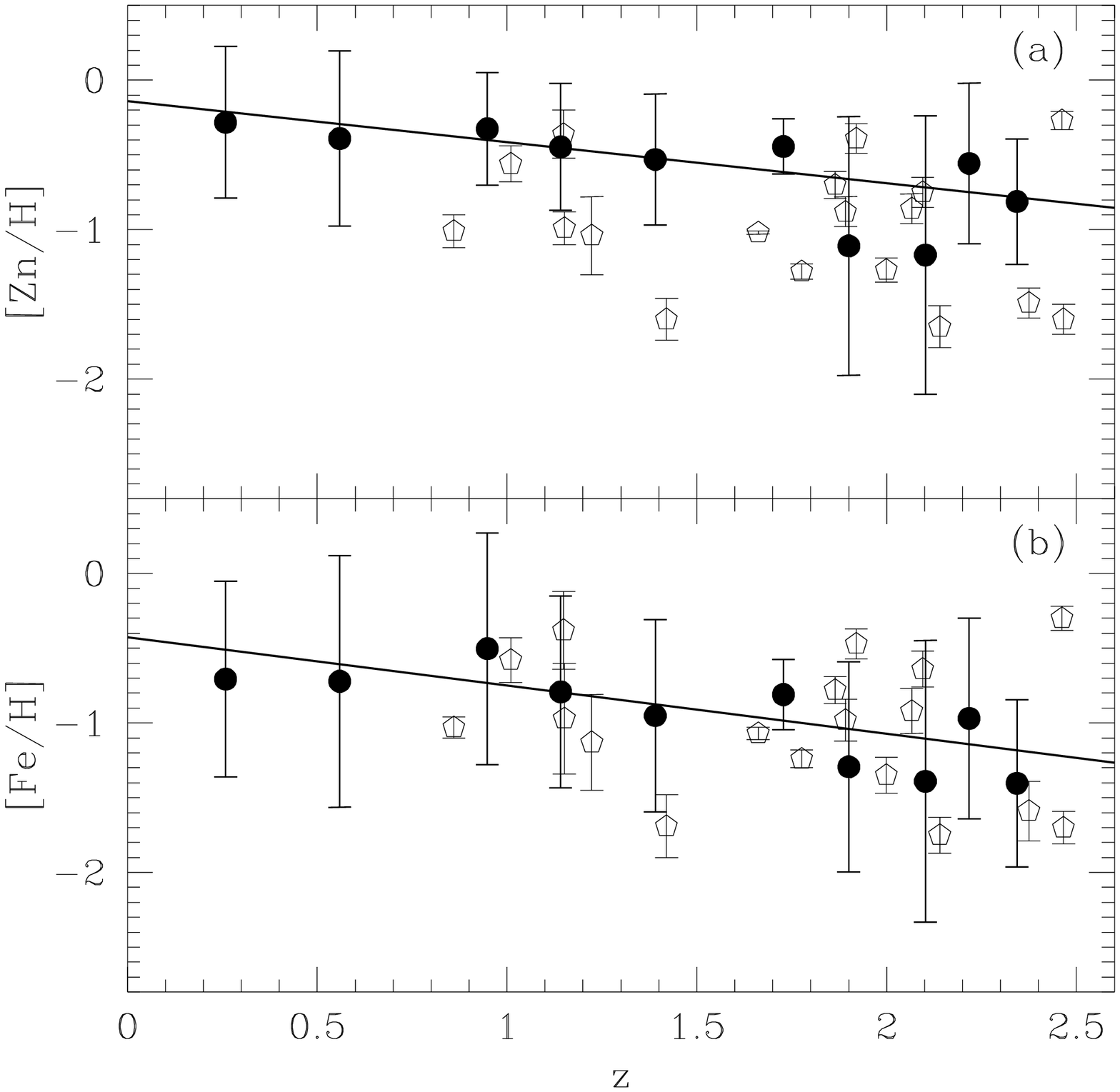}
\caption{$\langle {\rm [Zn/H]} \rangle$ (a) and $\langle {\rm [Fe/H]} \rangle$ (b) 
unweighted mean abundances for
the neutral hydrogen along the simulated 
LOSs with $N({\rm H_{I}}) > 10^{20} {\rm atoms \ cm^{-2}}$
 as a function of redshift
({\em filled circles}).  Error bars correspond to $1\sigma$ standard
deviation.
Solid lines are the least square linear regression for
 the whole sample of simulated DLAs.
We have included the observational data of Vladilo (2002) that 
incorporate dust-corrections ({\em open pentagons}). 
}
\label{znfe_z}
\end{figure}

\begin{figure}
\includegraphics[width=84mm]{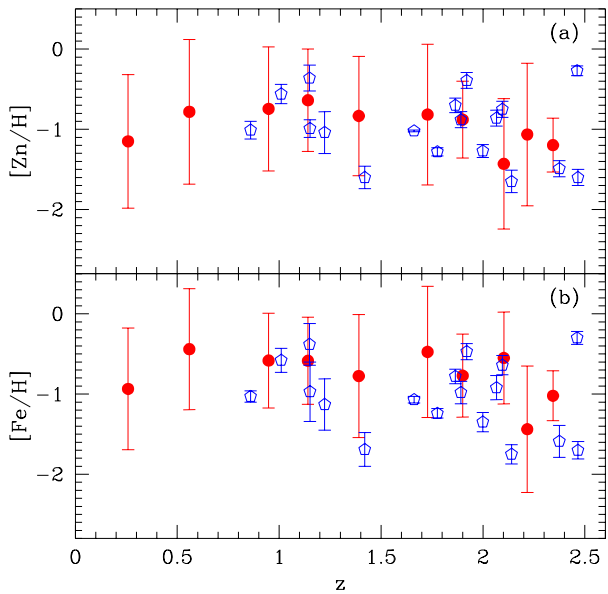}
\caption{$\langle {\rm [Zn/H]} \rangle$ (a) and $\langle {\rm [Fe/H]} \rangle$ (b) 
unweighted mean abundances for
the stellar population associated to the simulated DLAs shown in 
Fig.1 Abundances for dust corrected observed DLAs (Vladilo 2002) have been included for
comparison ({\em open pentagons}). 
}

\label{znfe_z_stars}
\end{figure}

\begin{figure}
\includegraphics[width=84mm]{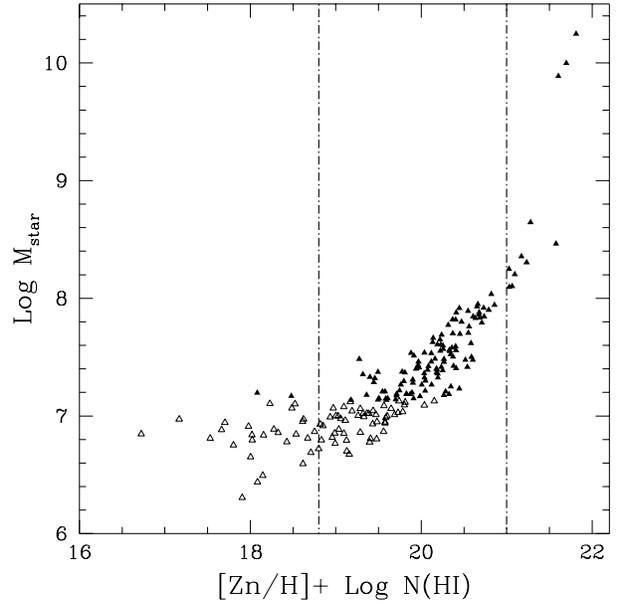}
\label{mstar_filter}
\caption{The stellar masses (in units of $M_{\odot}$)
 associated to the simulated DLAs versus the observational filter
$F$=[Zn/H] + log $N({\rm H_I})$, suggested by Boiss\'e et al. (1998),
for DLAs ($N({\rm H_I}) > 2 \times 10^{20}
\ {\rm atoms \ cm^{-2}}$; {\em filled triangles}) and sub-DLAs
 ($10^{19} <  N({\rm H_I})  < 2 \times 10^{20}
\ {\rm atoms \ cm^{-2}}$; {\em open triangles}). The dashed-dotted lines
depict the  region where observed DLAs lay: $18.8 < F < 21$.}
\end{figure}

\begin{figure}
\includegraphics[width=84mm]{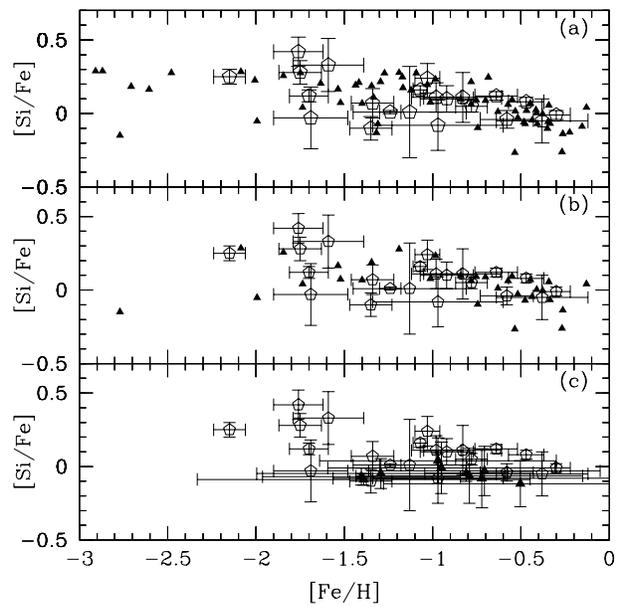}
\label{sife}
\caption{[Si/Fe] versus [Fe/H] relations obtained from simulations
({\em filled triangles}): 
a) each simulated DLA, 
b) each filtered simulated DLA and
c) unweighted means ${\langle{\rm [Si/Fe]}\rangle}$ and ${\langle{\rm [Fe/H]}\rangle}$ 
for neutral hydrogen along LOSs  
for all redshifts analysed.
We have included the dust-corrected observations of Vladilo (2002; {\em open
pentagons}).
}
\end{figure} 



\begin{figure}
\includegraphics[width=84mm]{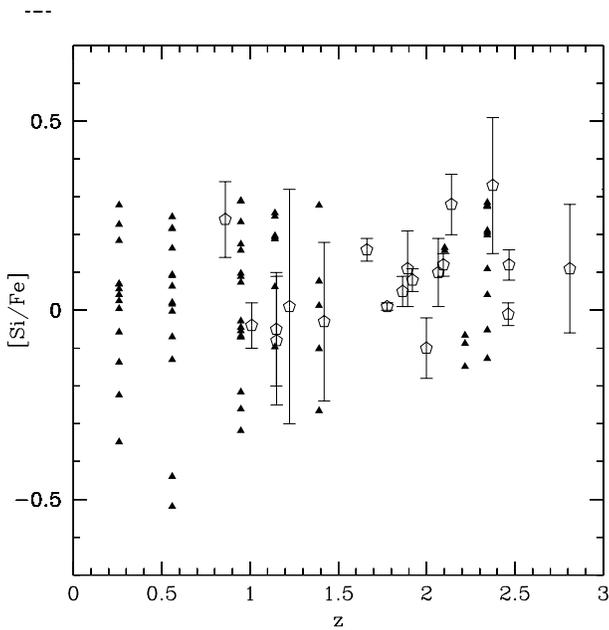}
\label{sifefeh}
\caption{ [Si/Fe] versus redshift for  the neutral hydrogen  
in DLAs ({\em filled triangles}).
Dust corrected observational data have been included for
comparison
(Valdilo 2002, {\em open pentagons}).}
\end{figure} 

\begin{figure}
\includegraphics[width=84mm]{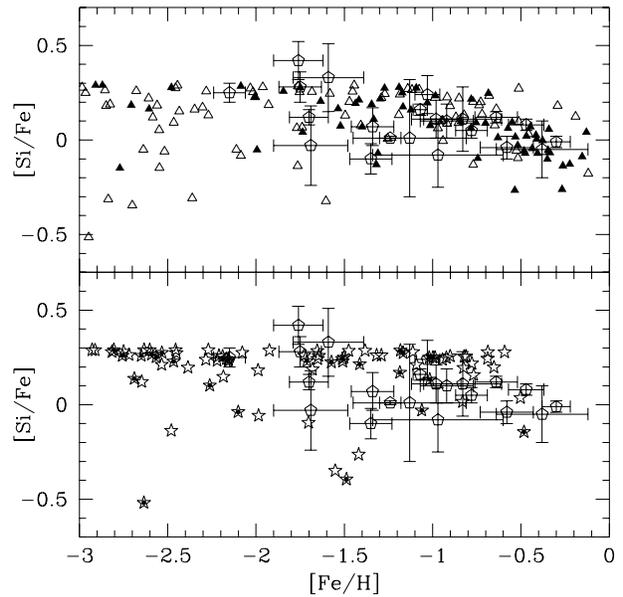}
\label{sifefeh}
\caption{ [Si/Fe] versus [Fe/H] for  the neutral hydrogen  
(a) and 
 the stellar populations (b)  
in DLAs ({\em filled triangles and dotted stars}, respectively)
 and   sub-DLAs ({\em open triangles and stars}, respectively).
Dust corrected observational data have been included for
comparison
(Valdilo 2002, {\em open pentagons}).}
\end{figure}

\begin{figure}
\includegraphics[width=84mm]{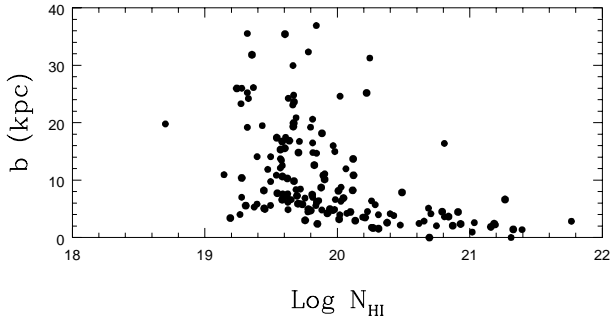}
\label{nhimpact}
\caption{Impact parameter versus ${\rm H_{I}}$ column density for LOSs drawn
through galaxy-like objects in the redshift range $0.26 < z < 2.35$. }
\end{figure}

\begin{figure}
\label{impactomb}
\includegraphics[width=84mm]{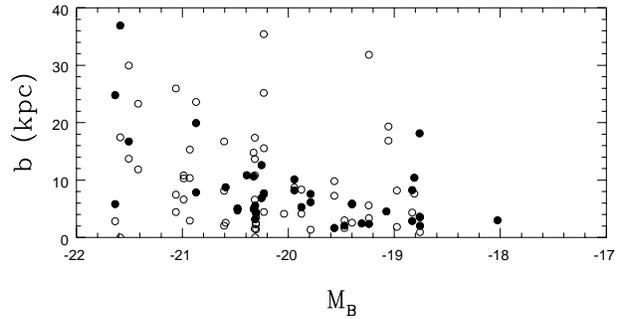}
\caption{ Impact parameter $b$ (kpc) as a function of the blue absolute
magnitude $M_{B}$ for simulated DLAs and sub-DLAs at $z<1$.
We have distinguished between those satisfying the observational
constrain $18.8 <$ log $N({\rm H_{I}})$ + [Zn/H] $<21$ ({\em filled circles})
and those outside this observational window ({\em open circles}).
}
\end{figure}

\begin{figure}
\label{lumz}
\includegraphics[width=84mm]{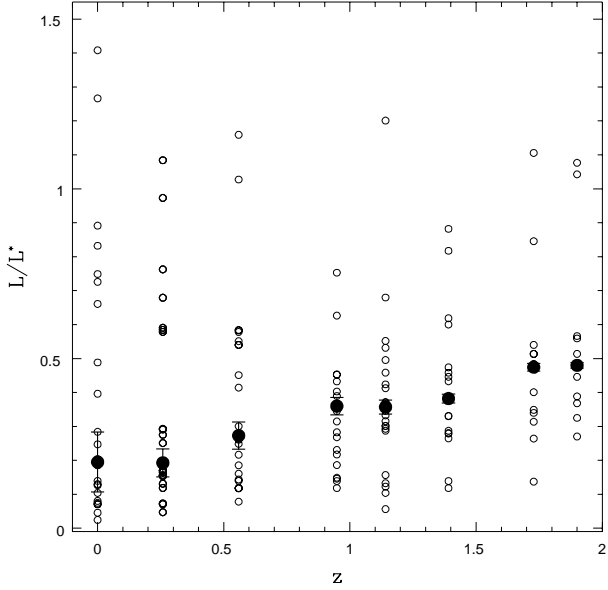}
\caption{ The median blue luminosity of building blocks that
host the  
$N({\rm H_{I}})$ absorptions
 normalized to $L^{\star}_{B}$ at $z=0$ as a function of {z} 
({\em
filled circles}). Bootstrap errors are shown. We have also included
the individual ratios  ({\em open circles}).}

\end{figure}

\begin{figure}
\label{rcutgas}
\includegraphics[width=84mm]{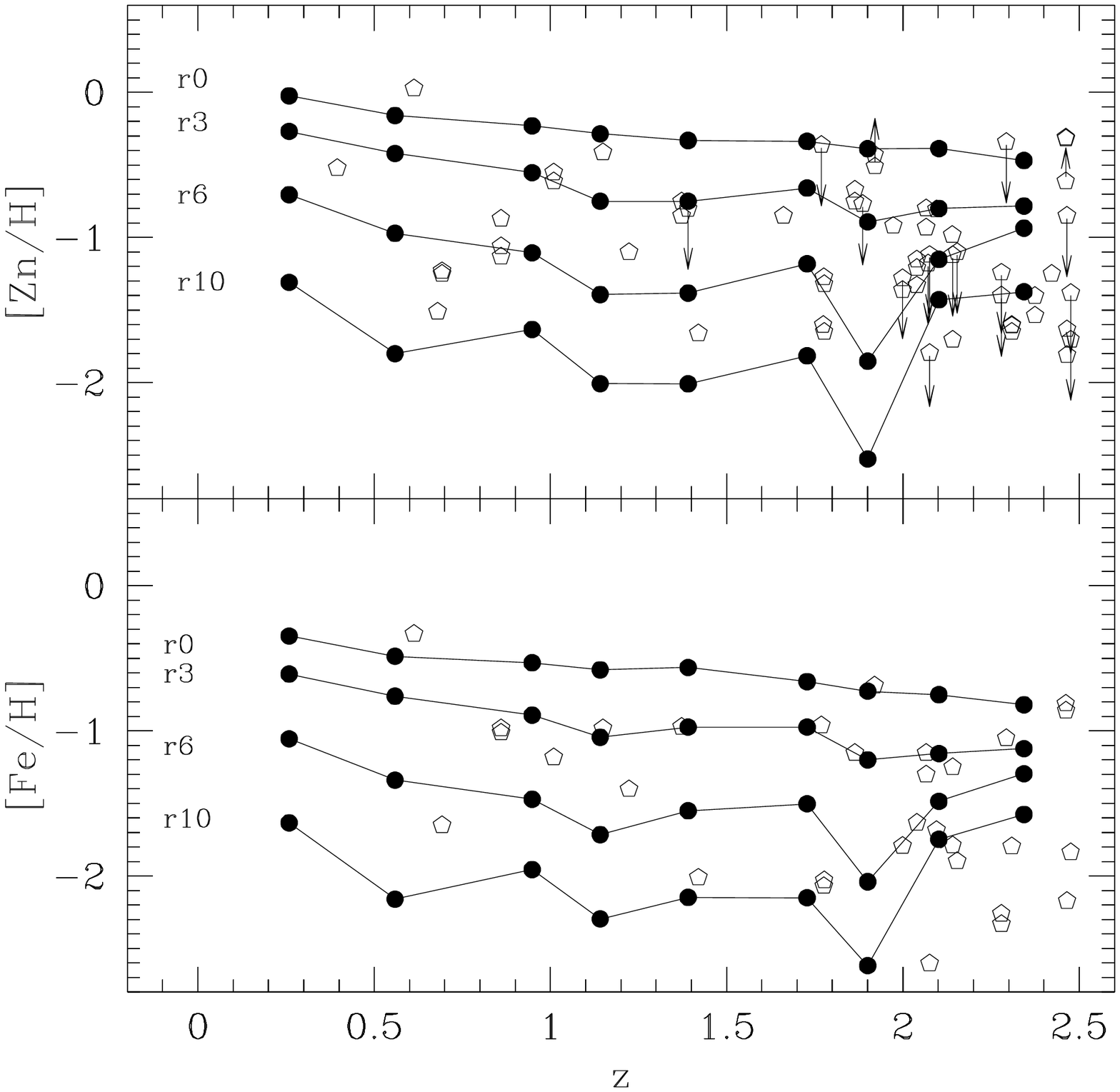}
\caption{$\langle {\rm [Zn/H]} \rangle$ and $\langle {\rm [Fe/H]} \rangle$ 
unweighted mean abundances as a function of redshift for
 gas in GLOs ({\em filled circles})
further than a certain radius from the centre
of the simulated galactic objects.
Connected symbols from the top to the bottom
represent inner cut-off radius
of $r0=0, r3=3, r6=6$ and $r10=10$ kpc, respectively.
Observational data is represented by {\em open pentagons} (Vladilo 2002)}.
\end{figure}

\begin{figure}
\label{rcutstars}
\includegraphics[width=84mm]{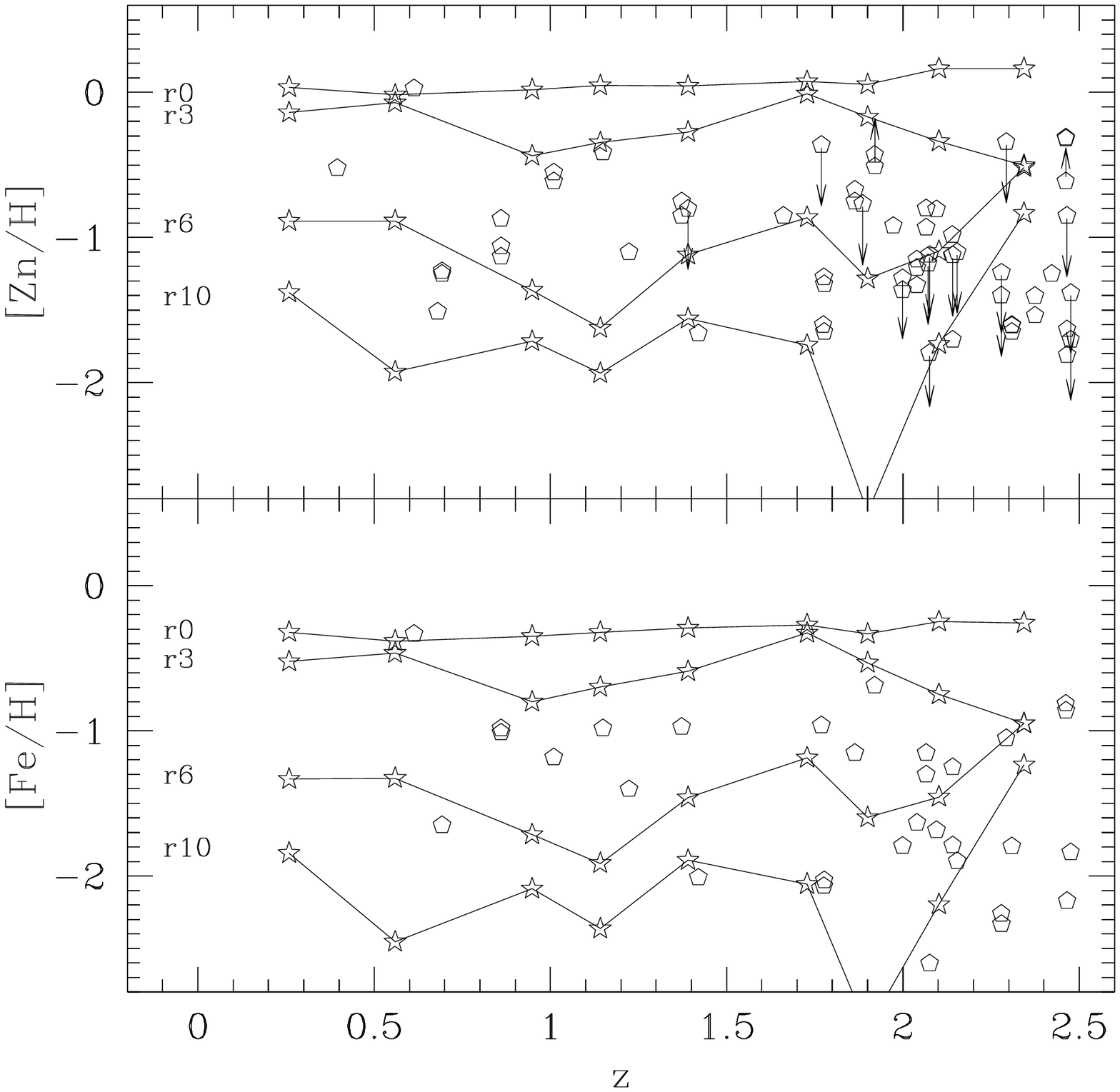}
\caption{$\langle {\rm [Zn/H]} \rangle$ and $\langle {\rm [Fe/H]} \rangle$ 
unweighted mean abundances as a function 
of redshift for
stars in GLOs ({\em empty stars})
further than a certain radius from the centre
of the simulated galactic objects.
Connected symbols from the top to the bottom
represent inner cut-off radius
of $r0=0, r3=3, r6=6$ and $r10=10$ kpc, respectively.
Observational data is represented by {\em open pentagons}}
\end{figure}


\begin{table*}
\centering
\begin{minipage}{140mm}
\caption{Chemical Evolution: Linear Regressions (${\rm LR} \equiv d{\rm logX}/dz$) 
through observed (${\rm LR^{\rm obs}}$) and simulated (${\rm LR^{\rm sim}}$) DLAs (bootstrap errors given), 
and standard dispersions of these samples in redshift bins $z_{\rm low}$ and $z_{\rm inter}$. For observations, we have included the corresponding
values for the dust-corrected samples (${\rm LR^{\rm obs}_{\rm dust}}$), while for simulated DLAs
the filtered (${\rm LR ^{\rm sim}_{\rm fil}}$) estimations are also displayed.}
\begin{tabular}{@{}ccccccccc@{}}
\hline
X &       ${\rm LR}^{\rm obs}$ &  ${\rm LR}^{\rm obs}_{\rm dust}$ & $\sigma^{\rm obs}_{\rm z_{low}}$ & $\sigma^{\rm obs}_{\rm z_{inter}}$ &  ${\rm LR}^{\rm sim}$ & ${\rm LR}^{\rm sim}_{\rm fil}$        & $\sigma^{\rm sim}_{\rm z_{low}}$ & $\sigma^{\rm sim}_{\rm z_{inter}}$\\
\hline
[Fe/H] &  $-0.17 \pm 0.11$ & $-0.31 \pm 0.14$ & 0.42                       & 0.68                        & $-0.36 \pm 0.05 $ & $-0.36 \pm 0.04$&    0.59           & 0.58               \\

[Zn/H] &  $-0.28 \pm 0.13$ & $-0.28 \pm 0.13 $ & 0.39                       & 0.65                        &$-0.26 \pm 0.03 $& $-0.26 \pm 0.03$  & 0.33              & 0.44                \\

[Si/H] &  $-0.19 \pm 0.24 $& $-0.22 \pm 0.24 $ & 0.15                       &0.12                         &$-0.29 \pm 0.06$ &$ -0.32 \pm 0.03$  &                0.45   &         0.48            \\

[Si/Fe]&  $-0.02 \pm 0.09$ & $\quad 0.09 \pm 0.08$ & 0.10  &0.31   &$\quad 0.05  \pm 0.01 $ & $\quad 0.04 \pm 0.01$ &   0.18                &   0.13                  \\

\hline

\end{tabular}
\end{minipage}
\end{table*}

\begin{table*}
\centering
\begin{minipage}{140mm}
\caption{Intrinsic Evolution: difference between unweighted means estimated at redshift bins 
$z_{\rm low}$ and $z_{\rm inter}$ for both observed and simulated DLAs
(bootstrap errors given).} 
\begin{tabular}{@{}ccccc@{}}
\hline

[X/Y]  & $\Delta {\langle {\rm [X/Y]}^{\rm obs} \rangle}_{\rm u}$ &  $\Delta {\langle {\rm [X/Y]}^{\rm obs}_{\rm dust} \rangle}_{\rm u}$ &  $\Delta {\langle {\rm [X/Y]}^{\rm sim} \rangle}_{\rm u}$ & $\Delta {\langle {\rm [X/Y]}^{\rm sim}_{\rm fil} \rangle}_{\rm u}$\\
\hline

[Fe/H] & $-0.09\pm 0.18 $ & $-0.06 \pm 0.24$  &  $-0.33 \pm 0.11$& $-0.21 \pm 0.08$        \\

[Zn/H] & $-0.05\pm 0.20 $ & $-0.05 \pm 0.20$  &  $-0.23 \pm 0.09$ & $-0.03 \pm 0.09$        \\

[Si/H] & $-0.03 \pm 0.20$& $-0.02 \pm 0.20$   & $-0.29 \pm 0.12 $ & $-0.19 \pm 0.08$        \\

[Si/Fe]& $-0.12 \pm 0.08$ & $ -0.08\pm 0.05$  & $-0.02 \pm 0.09$ & $\quad 0.02 \pm 0.02$          \\

\hline

\end{tabular}
\end{minipage}
\end{table*}

\begin{table*}
\centering
\begin{minipage}{140mm}
\caption{Global Evolution: difference between ${\rm H_I}$ mass weighted means 
estimated at redshift bins 
$z_{\rm low}$ and $z_{\rm inter}$ for both observed and simulated DLAs 
(bootstrap errors given).} 
\begin{tabular}{@{}ccccc@{}}
\hline
[X/Y]  & $\Delta {\langle {\rm [X/Y]}^{\rm obs} \rangle}_{\rm w}$ &  $\Delta {\langle {\rm [X/Y]}^{\rm obs}_{\rm dust} \rangle}_{\rm w}$ &  $\Delta {\langle {\rm [X/Y]}^{\rm sim} \rangle}_{\rm w}$ & $\Delta {\langle {\rm [X/Y]}^{\rm sim}_{\rm fil} \rangle}_{\rm w}$\\
\hline
[Fe/H] & $-0.12 \pm 0.08$ & $-0.31 \pm 0.35$ &  $-0.85 \pm 0.48$ & $-0.24 \pm 0.21$      \\

[Zn/H] & $-0.30\pm 0.34$ & $-0.30 \pm 0.34$   &  $-0.51 \pm 0.39 $ & $-0.14 \pm 0.04$      \\

[Si/H] & $-0.04 \pm 0.23$ & $-0.19 \pm 0.37$  & $-0.72 \pm 0.59 $ & $-0.27 \pm 0.09$        \\

[Si/Fe]& $-0.29 \pm 0.10$ & $-0.04 \pm 0.08$ & $-0.32 \pm 0.12$ & $-0.18\pm 0.05$        \\

\hline

\end{tabular}
\end{minipage}
\end{table*}

\end{document}